\documentclass[pra,showpacs,preprintnumbers,amsmath,amssymb]{revtex4}
\usepackage{amssymb,amsmath}
\usepackage{color}
\usepackage{graphicx}
\usepackage{dcolumn}
\usepackage{bm}
\usepackage{latexsym,epsfig}

\begin{document}

\title{The Inhomogeneous Extended Bose-Hubbard Model: A Mean-Field Theory}

\author{Jamshid Moradi Kurdestany}
\email{jamshid@physics.iisc.ernet.in, jmkurdestany@gmail.com}
\affiliation{Department of Physics, Indian Institute of Science,
Bangalore 560 012, India}

\author{Ramesh V. Pai}
 \email{rvpai@unigoa.ac.in}
 \affiliation{Department of Physics,
Goa University, Taleigao Plateau, Goa 403 206, India}
\author{Rahul Pandit}
\email{rahul@physics.iisc.ernet.in} \altaffiliation[Also
at~]{Jawaharlal Nehru Centre For Advanced Scientific Research,
Jakkur, Bangalore, India} \affiliation{Centre for Condensed Matter
Theory, Department of Physics, Indian Institute of Science,
Bangalore 560012, India.}

\date{\today}
\begin{abstract}
We develop an inhomogeneous mean-field theory for the extended Bose-Hubbard
model with a quadratic, confining potential. In the absence of this
potential, our mean-field theory yields the phase diagram of the homogeneous
extended Bose-Hubbard model. This phase diagram shows a superfluid (SF) phase
and lobes of Mott-insulator (MI), density-wave (DW), and supersolid (SS)
phases in the plane of the chemical potential $\mu$ and on-site repulsion
$U$; we present phase diagrams for representative values of $V$, the
repulsive energy for bosons on nearest-neighbor sites. We demonstrate that,
when the confining potential is present, superfluid and density-wave order
parameters are nonuniform; in particular, we obtain, for a few representative
values of parameters, spherical shells of SF, MI, DW, and SS phases. We
explore the implications of our study for experiments on cold-atom dipolar
condensates in optical lattices in a confining potential
\end{abstract}

 \maketitle

\section{Introduction}
\label{sec:intro}

Experimental studies of quantum phase
transitions~\cite{rmp,advphys,jaksch,greiner} in systems of cold atoms in
traps, with an imposed optical lattice, have led to a renewal of interest
in theoretical studies of lattice models of interacting
bosons~\cite{fisher,mc,sheshadri}. Examples of such transitions include one
from a superfluid (SF) to a bosonic Mott-insulator (MI).  This transition was
predicted by mean-field theories, such as those of
Refs.~\cite{fisher,sheshadri}, and obtained in Monte-Carlo
simulations~\cite{mc} of the Bose-Hubbard model before it was realized
experimentally.

In addition to the optical-lattice potential, a confining
potential, most often quadratic, is present in all cold-atom experiments.
This leads to inhomogeneities in the phases that are obtained:
simulations~\cite{wessel,svistunov} of the Bose-Hubbard model, with such a
confining potential, and experiments~\cite{bloch,campbel} on interacting
bosons in optical lattices, with a confining potential, have both seen
alternating shells of SF and MI regions.

Mean-field theories for the Bose-Hubbard model were first developed for the
homogeneous case~\cite{fisher,sheshadri,rvpai08}. These theories were then
extended to the inhomogeneous case~\cite{sheshprl} to develop an
understanding of the Bose-glass phase in the disordered Bose-Hubbard model.
In recent work~\cite{rvpaiinhom11} we have shown how the effects of such a
confining potential can be treated, at the level of mean-field theory, as was
done in the Bose-glass case~\cite{sheshprl}; in particular, we have provided
a natural framework for understanding alternating SF and MI shells, mentioned
above. Here we extend this inhomogeneous mean-field theory to account for the
different types of phases, SF, MI, density-wave (DW), and supersolid (SS),
which can occur in the extended Bose-Hubbard model~\cite{kovrizhin05}.

The principal motivation for undertaking such a study of the
extended Bose-Hubbard model is provided by the experiments that
have obtained a dipolar condensate of $^{52}{\rm Cr}$
atoms~\cite{werner05}. To understand these
experiments we must study lattice models of bosons with
long-range interactions~\cite{goral02} and
not merely the Bose-Hubbard model with a repulsive
interaction between bosons on the same lattice site. The simplest
model that goes beyond such onsite interactions is the extended
Bose-Hubbard model, which allows for repulsive interactions
between bosons on nearest-neighbor sites and the
aforementioned onsite interaction. In addition to SF and MI
phases of the Bose-Hubbard model, this extended model can have
a density wave (DW) phase, in which the mean density of bosons is
different on the two sublattices of the hypercubic lattices we
consider, and a super-solid (SS) phase (see, e.g.,
Refs.~\cite{kovrizhin05,andreev69,kim04}).

Before we present the details of our study, we summarize our principal
results. We first develop a mean-field theory for the homogeneous, extended
Bose-Hubbard model by developing on the work of our group on Bose-Hubbard
models for the spinless and spin-1 cases~\cite{sheshadri,rvpai08}; this
yields the SF, MI, DW, and SS phases and the transitions between them, which
have been studied by a Gutzwiller-type approximation~\cite{kovrizhin05} that
is akin to, but not the same as, our mean-field theory. We then develop an
inhomogeneous mean-field theory for the inhomogeneous extended,
Bose-Hubbard model by generalizing our inhomogeneous mean-field theory for
the Bose-Hubbard model~\cite{rvpaiinhom11}.  In particular, when we use a
quadratic confining potential in three dimensions (3D), our theory yields
inhomogeneous phases with spherical shells of SF, MI, DW, and SS states. The
precise way in which these phases alternate depends on the parameters of the
model; we study a few illustrative cases explicitly for which we present
order-parameter profiles and their Fourier transforms. We also discuss the
experimental implications of our work.

The remaining part of this paper is organized as follows. In
Sec.~\ref{sec:modelmft} we introduce the inhomogeneous extended
Bose-Hubbard model and then develop an inhomogeneous mean-field
theory for it. In Sec.~\ref{sec:results} we present the results
of our mean-field theory. Section~\ref{sec:conclusions} contains
concluding remarks; here we give a brief comparison of our work
with earlier studies and we explore the experimental implications
of our study.

\section{Model and Mean-Field Theory}
\label{sec:modelmft}

We study the inhomogeneous, extended Bose-Hubbard model that is defined by
the Hamiltonian
\begin{eqnarray} \frac{{\cal H}}{zt} &=& -\frac{1}{z}
\sum_{<i,j>} (a_{i}^{\dagger} a_{j} + h.c)+\frac{1}{2}\frac{U}{zt} \sum_{i}
{\hat n}_{i} ({\hat n}_{i} -1) \nonumber \\
&+&\frac{V}{zt} \sum_{<i,j>}{\hat n}_i{\hat n}_{j}- \frac{1}{zt}
\sum_i\mu_i {\hat n}_i ,
\label{eq:ebhmodel}
\end{eqnarray}
where $t$ is the amplitude for a boson to hop from site $i$ to its
nearest-neighbor site $j$, $z$ is the nearest-neighbor coordination number,
$<i,j>$ are nearest-neighbor pairs of sites, $h.c.$ denotes the Hermitian
conjugate, $a_i^{\dag}, \, a_i$, and ${\hat n}_i \equiv a_i^{\dag} a_i$ are,
respectively, boson creation, annihilation, and number operators at the site
$i$, the repulsive potential between bosons on the same site is $U$, the
chemical potential $\mu_i$ controls the number of bosons at the site $i$, and
$V$ is the repulsive interaction between bosons on nearest-neighbor sites. In
the inhomogeneous case, the chemical potential is $\mu_i \equiv \mu-V_T
R^2_i$, where $\mu$ is the uniform part of the chemical potential, $V_T$ the
strength of the harmonic confining potential, $R^2_i\equiv \sum_{n=1}^d
X^2_n(i)$, where $X_n(i),\, 1\leq n \leq d$, are the Cartesian coordinates of
the site $i$ and  $d$ is dimension of the hypercubic lattice (we study $d=3$
explicitly); the origin is chosen to be at the center of this lattice.
Clearly, if we set $V = 0$, we obtain the inhomogeneous Bose-Hubbard model of
Ref.~\cite{rvpaiinhom11}, which we follow in our mean-field treatment below.
In this paper, we set $zt=1$, i.e., we measure all energies in units of
$zt$.

If $t=0$ and $V_T=0$, the model~(\ref{eq:ebhmodel}) exhibits (a) MI phases,
which have an integral number of bosons per site, or (b) DW phases, in which
bosons preferentially occupy one of the sublattices, say ${\mathcal A}$, of
the bipartite, hypercubic lattices we consider; the MI phases are favored at
large values of $U$ whereas the DW phases appear if $V$ is large.  A variety
of DW phases are possible; we denote them by DW M/2; here M is the number of
atoms per unit cell and 2 denotes that the unit cell is doubled, i.e., the
length of its side is 2. For example, when $t=0$ and $V_T = 0$, the DW 1/2
phase has 1 boson on sublattice ${\mathcal A}$ and none on sublattice
${\mathcal B}$ (or vice versa); in DW 3/2 there is 1 boson on sublattice
${\mathcal A}$ and 2 on sublattice ${\mathcal B}$ (or vice versa).

If $V_T=0$ but $t\neq 0$, SF or SS phases can be stabilised because the
bosons can hop through the lattice. Nonuniform states appear when we
allow $V_T \neq 0$ as we show below via our inhomogeneous mean-field theory.

We now generalize the intuitively appealing mean-field theory of
Ref.~\cite{sheshadri}, which has been developed for the homogeneous
Bose-Hubbard model and then extended to the inhomogeneous case in
Refs.~\cite{sheshprl,rvpaiinhom11}. Our generalization introduces
order parameters that are capable of distinguishing between DW, SS,
SF, and MI phases.  Conventional mean-field theories introduce a
decoupling scheme that reduces a model with interacting bosons or
fermions to an effective, noninteracting problem, which can be
solved easily because the effective Hamiltonian is quadratic in
boson or fermion operators. By contrast, the mean-field theories of
Refs.~\cite{sheshadri,rvpaiinhom11}, for the case $V=0$, decouple
the hopping term in Eq.~(\ref{eq:ebhmodel}), which is quadratic in
boson operators, to obtain an effective, one-site Hamiltonian, which
can be diagonalized numerically. To generalize this to the case $V >
0$, we have to decouple the number operators in the extended
Bose-Hubbard term in Eq.~(\ref{eq:ebhmodel}).  In particular, we
decouple the first and third terms of Eq.~(\ref{eq:ebhmodel}) to
obtain an effective one-site problem, which neglects quadratic
deviations from equilibrium values (denoted by angular brackets).
The two approximations we use are as follows:
\begin{eqnarray} a^{\dagger}_{i}a_{j} &\simeq& \langle a^{\dagger}_{i}\rangle
a_{j} +a^{\dagger}_{i}\langle a_{j}\rangle -\langle a^{\dagger}_{i}\rangle
\langle a_{j}\rangle ; \nonumber \\
{\hat n}_{i}{\hat n}_{j} &\simeq& \langle
{\hat n}_{i}\rangle {\hat n}_{j} +{\hat n}_{i}\langle {\hat n}_{j}\rangle
-\langle {\hat n}_{i}\rangle \langle {\hat n}_{j}\rangle;
\label{eq:decoup}
\end{eqnarray}
here the superfluid order parameter and the local density for
the site $i$ are, respectively, $\psi_{i}\equiv \langle a_{i}\rangle$ and
$\rho_{i} \equiv \langle {\hat n}_{i}\rangle $, respectively.  The
approximation~(\ref{eq:decoup}) can now be used to write the
Hamiltonian~(\ref{eq:ebhmodel}) as a sum over single-site, mean-field
Hamiltonians ${\cal{H}}_i^{MF}$ as follows:
\begin{eqnarray}
{\cal{H}}^{MF} &\equiv& \sum_{i} {\cal{H}}_i^{MF} , \nonumber \\
\frac{{\cal{H}}_i^{MF}}{zt} &\equiv&
\frac{1}{2}\frac{U}{zt} {\hat{n}}_i({\hat{n}}_i-1) -\frac{\mu_i}{zt}
{\hat{n}}_i -(\phi_ia^{\dagger}_{i}+\phi_i^*a_{i}) \nonumber \\
&+&\frac{1}{2}(\psi_i^* \phi_i+\psi_i \phi_i^*)+
{\frac{V}{t}}({\hat{n}}_i{\bar{\rho}}_i-{\rho_i}{\bar{\rho}}_i),
\label{eq:mfham}
\end{eqnarray}
where the superscript $MF$ stands for mean
field, and $\phi_i\equiv \frac{1}{z}\sum_{\delta} \psi_{i+\delta}$ ,
${\bar{\rho}}_i\equiv \frac{1}{z}\sum_{\delta} \rho_{i+\delta}$, and $\delta$
labels the $z$ nearest neighbors of the site $i$. This form of the
single-site, mean-field Hamiltonian is suitable for the inhomogeneous case
with $V_T > 0$.

For the homogeneous case, we note that the hypercubic lattices we consider are
bipartite, i.e., they can be divided into two sublattices ${\mathcal A}$ and
${\mathcal B}$. Each site on the ${\mathcal A}$ (${\mathcal B}$) sublattice
has $z$ nearest neighbors each one of which belongs to the ${\mathcal B}$
(${\mathcal A}$) sublattice. Thus, if $V_T = 0$, $\psi_{i}=\psi_A$ and
$\rho_{i}=\rho_A$ if $i\in {\mathcal A}$ and $\psi_{i}=\psi_B$ and
$\rho_{i}=\rho_B$ if $i\in {\mathcal B}$, whereas $\phi_{i}=\psi_B$ and
${\bar \rho}_{i}=\rho_B$ if $i\in {\mathcal A}$ and $\phi_{i}=\psi_A$ and
${\bar \rho}_{i}=\rho_A$ if $i\in {\mathcal B}$.  If we require chemical
potentials that are conjugate to $\rho_A$ and $\rho_B$, respectively, we can
introduce $\mu_{i}=\mu_A$ if $i\in {\mathcal A}$ and $\mu_{i}=\mu_B$ if $i\in
{\mathcal B}$; similarly, we can define creation, annihilation, and number
operators for each sublattice and hence write the
mean-field Hamiltonian~(\ref{eq:mfham}) as follows:
\begin{equation}
\cal{H}_{AB}^{MF} \equiv \cal{H}_{A}^{MF}+\cal{H}_{B}^{MF};
\label{eq:hab1}
\end{equation}

\begin{eqnarray}
\frac{{\cal{H}}_A^{MF}}{zt} &\equiv&
-(a_{A}\psi_B^*+a^{\dagger}_{A}\psi_B)+
\frac{1}{2}(\psi_A\psi_B^*+\psi_A^*\psi_B)
\nonumber \\
&&+ \frac{V}{t}({\hat{n}}_A{\rho}_B-{\rho}_A{\rho}_B)+
\frac{1}{2}\frac{U}{zt}{{\hat{n}}_A}({{\hat{n}}_A}-1)-\frac{\mu_A}{zt}
{\hat{n}}_A; \label{eq:hab2}
\end{eqnarray}

\begin{eqnarray}
\frac{{\cal{H}}_B^{MF}}{zt} &\equiv&
-(a_{B}\psi_A^*+a^{\dagger}_{B}\psi_A)+
\frac{1}{2}(\psi_B\psi_A^*+\psi_B^*\psi_A)
\nonumber \\
&&+ \frac{V}{t}({\hat{n}}_B{\rho}_A-{\rho}_B{\rho}_A)+
\frac{1}{2}\frac{U}{zt}{\hat{n}}_B({\hat{n}}_B-1)-\frac{\mu_B}{zt}
{\hat{n}}_B. \label{eq:hab3}
\end{eqnarray}

If $V_T > 0$, we first obtain the matrix elements of
${\cal{H}}_i^{MF}$ in the onsite, occupation-number basis
$\{|n_i\rangle\}$, truncated in practice by choosing a finite value
for $n_{{max}}$, the total number of bosons per site, for a given
initial set of values for $\{\psi_i,\rho_i\}$. The smaller the
values of $U$ and $V$ and the larger the value of $\mu$ the larger
must be the value of $n_{max}$; for the values of $U, \, V$, and
$\mu$ we consider $n_{max}=6$ suffices; we have checked this in
representative cases by carrying out calculations with
$n_{\mbox{max}}=10$.  We then diagonalize this matrix, which depends
not just on $\psi_i$ and $\psi_{i+\delta}$, but also on $\rho_i$ and
$\rho_{i+\delta}$, to obtain the lowest energy and the corresponding
wave function, denoted, respectively, by
$E_g^i(\psi_i,\psi_{i+\delta};\rho_i,\rho_{i+\delta})$ and
$\Psi_g(\{\psi_i\},\{\rho_i\})$; from these we obtain the new order
parameters $\psi_i = \langle \Psi_g(\{\psi_i\},\{\rho_i\})\mid a_i
\mid \Psi_g(\{\psi_i\},\{\rho_i\}) \rangle$ and $\rho_i = \langle
\Psi_g(\{\psi_i\},\{\rho_i\})\mid {\hat{n}}_i \mid
\Psi_g(\{\psi_i\},\{\rho_i\}) \rangle$. We then use these new values
of $\psi_i$ and $\rho_i$ as inputs to reconstruct ${\cal{H}}_i^{MF}$
and repeat the diagonalization procedure until we achieve self
consistency of input and output values to obtain the equilibrium
values $\psi^{eq}_i$ and $\rho^{eq}_i$ (we suppress the superscript
$eq$ hereafter for notational convenience). [This self-consistency
procedure is equivalent to a minimization of the total energy
$E_g(\{\psi_i\},\{\rho_i\})\equiv\sum_i
E_g^i(\psi_i,\psi_{i+\delta};\rho_i,\rho_{i+\delta})$ with respect
to $\psi_i$ and $\rho_i$ .] Given the form of the confining
potential, the self-consistent solutions for $\{\psi_i,\rho_i\}$
must have spherical (circular) symmetry in the three-dimensional
(two-dimensional) case; we use this spherical symmetry in obtaining
the self-consistent solutions. If $V_T=0$, we only need the four
order parameters $\psi_A, \, \psi_B, \, \rho_A$, and $\rho_B$ so the
problem of finding self-consistency solutions is much simpler than
it is in the inhomogeneous case with $V_T > 0$. In principle,
$\{\psi_i\}$ can be complex, but we find, as in earlier
calculations~\cite{sheshadri,rvpai08,sheshprl,rvpaiinhom11}, that
the equilibrium solution is such that $\{\psi_i\}$ are real.

\section{Results}
\label{sec:results}

In this Section we present the results of the inhomogeneous
mean-field theory that we have developed in the previous Section
for the extended Bose-Hubbard model~(\ref{eq:ebhmodel}). We
expect that the onsite repulsion between bosons is stronger than
the repulsive interaction between bosons on nearest-neighbor
sites, so we restrict ourselves to $V < U$. We begin with phase
diagrams for the homogeneous case with $V_T = 0$. We then
investigate order-parameter profiles in the presence of
the confining potential, i.e., when $V_T > 0$.

\begin{figure*}[htbp]
\epsfig{file=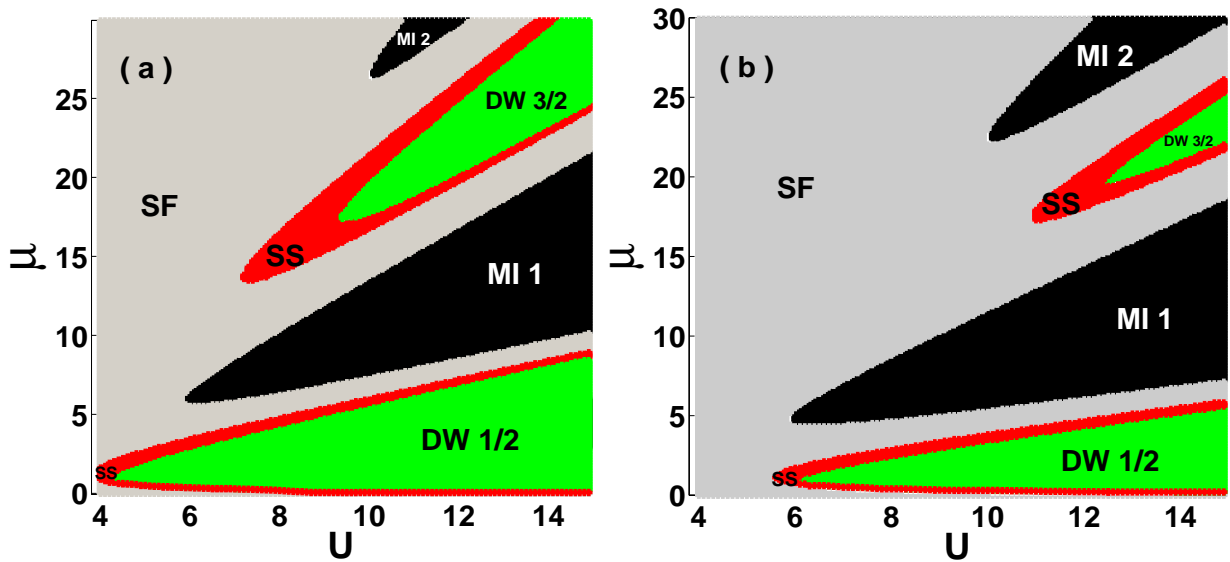,width=\linewidth,height=8cm} \caption{(Color
online) Phase diagrams in the $(\mu,\, U)$ plane for the extended
Bose-Hubbard model with $V_T = 0$ and (a) $V/U=0.6$ and (b)
$V/U=0.4$ showing SF (gray), SS (red), MI (black), and DW (green)
phases. MI1 and MI2 denote, respectively, MI phases with one and two
bosons per site; DW 1/2 and DW 3/2 are, respectively, DW phases with
one and three bosons per cubic unit cell with side 2; the basic
lattice spacing is taken to be 1; and $zt=1$.}
\label{fig:homphasediag}
\end{figure*}

In Figs.~\ref{fig:homphasediag} (a) and (b), we present phase
diagrams in the $(\mu,\, U)$ plane for the extended Bose-Hubbard
model~(\ref{eq:ebhmodel}), with $V_T = 0$ and (a) $V/U=0.6$ and (b)
$V/U=0.4$, with SF (gray), SS (red), MI (black), and DW (green)
phases; the MI phases MI1 and MI2 have, respectively, one and two
bosons per site; and DW 1/2 and DW 3/2 are, respectively, DW phases
with one and three bosons per unit cell with side $2$; we take the
lattice spacing of the underlying hypercubic lattice to be $1$. The
SF phase is favored at small values of $U$. If we hold $\mu$ fixed
at low values and increase $U$, the system first undergoes a
transition to an SS phase and then to the DW 1/2 phase. The lobe of
the MI1 phase appears above the DW 1/2 lobe and the encompassing
sliver of the SS phase; the next few DW and MI lobes appear as shown
in Figs.~\ref{fig:homphasediag} (a) and (b). Note that    the red
slivers of the SS phases encompass the DW lobes completely. Furthermore, the DW and SS phases grow at the expense of
the SF and MI phases as $V/U$ increases, as we expect for the
extended Bose-Hubbard model~(\ref{eq:ebhmodel}). The phase diagrams
of Figs.~\ref{fig:homphasediag} (a) and (b) are qualitatively
similar to those obtained by a Gutzwiller approximation in
Ref.~\cite{kovrizhin05}.

\begin{figure*}[htbp]
\epsfig{file=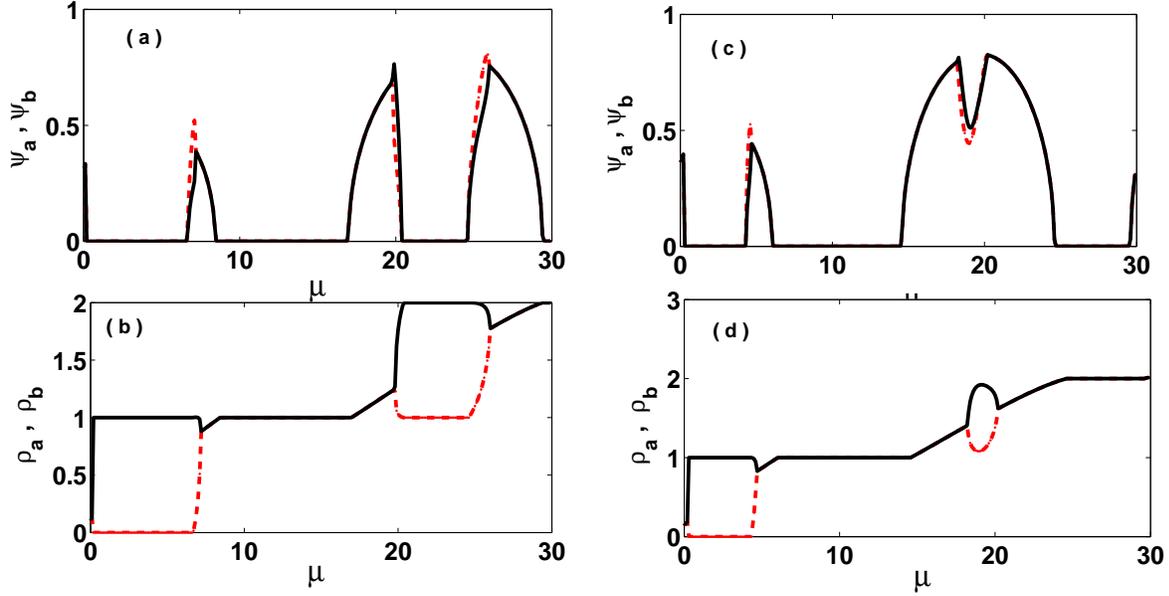,width=\linewidth,height=8cm} \caption{(Color
online) Plots of the superfluid order parameters $\psi_{a}$ (red
dashed line) and $\psi_{b}$ (black full line), on sublattices
${\mathcal A}$ and ${\mathcal B}$, versus $\mu/(zt)$ for $U/(zt)=12$ , $V_T=0$
and (a) $V/U=0.6$ and (c) $V/U=0.4$. Plots of the density-wave order
parameters $\rho_{a}$ (red dashed line) and $\rho_{b}$ (black full
line), on sublattices ${\mathcal A}$ and ${\mathcal B}$, versus
$\mu/(zt)$ for $U/(zt)=12$ and (b) $V/U=0.6$ and (d) $V/U=0.4$.}
\label{fig:homplots}
\end{figure*}

We obtain these phase diagrams by monitoring the dependence of the SF and DW
order parameters on $U, \, V$, and $\mu$.  In
Fig.~\ref{fig:homplots} we show representative plots of $\psi_{a}$ (red
dashed line) and $\psi_{b}$ (black full line), on sublattices ${\mathcal A}$
and ${\mathcal B}$, respectively, versus $\mu$ for $U=12$, $V_T = 0$, $zt=1$, and (a)
$V/U=0.6$ and (b) $V/U=0.4$. We also show representative plots of
$\rho_{a}$ (red dashed line) and $\rho_{b}$ (black full line), on
sublattices ${\mathcal A}$ and ${\mathcal B}$, respectively, versus $\mu$
for $U=12$, $zt=1$, and (c) $V/U=0.6$ and (d) $V/U=0.4$. We can distinguish
these phases from each other by noting the following: in the SF phase $\psi_A
= \psi_B > 0$ and $\rho_A = \rho_B$; in the SS phases $\psi_A, \psi_B > 0$,
$\psi_A \neq \psi_B$, and $\rho_A \neq \rho_B$; in the DW phases $\psi_A
= \psi_B = 0$ but $\rho_A \neq \rho_B$; in the MI phases $\psi_A = \psi_B =
0$ and $\rho_A = \rho_B = m$, a positive integer.

We now use the inhomogeneous mean-field theory, which we have developed in
the previous Section, to obtain alternating spherical shells of MI, SF, DW,
and SS phases in the 3D, extended Bose-Hubbard model~(\ref{eq:ebhmodel}) with
a quadratic confining potential. We do this by obtaining the order-parameter
profiles $\{\psi_i,\rho_i\}$ and also by obtaining in-trap density
distributions of bosons at different values of $U, \, V$, and $\mu$.

\begin{figure}[htbp]
\epsfig{file=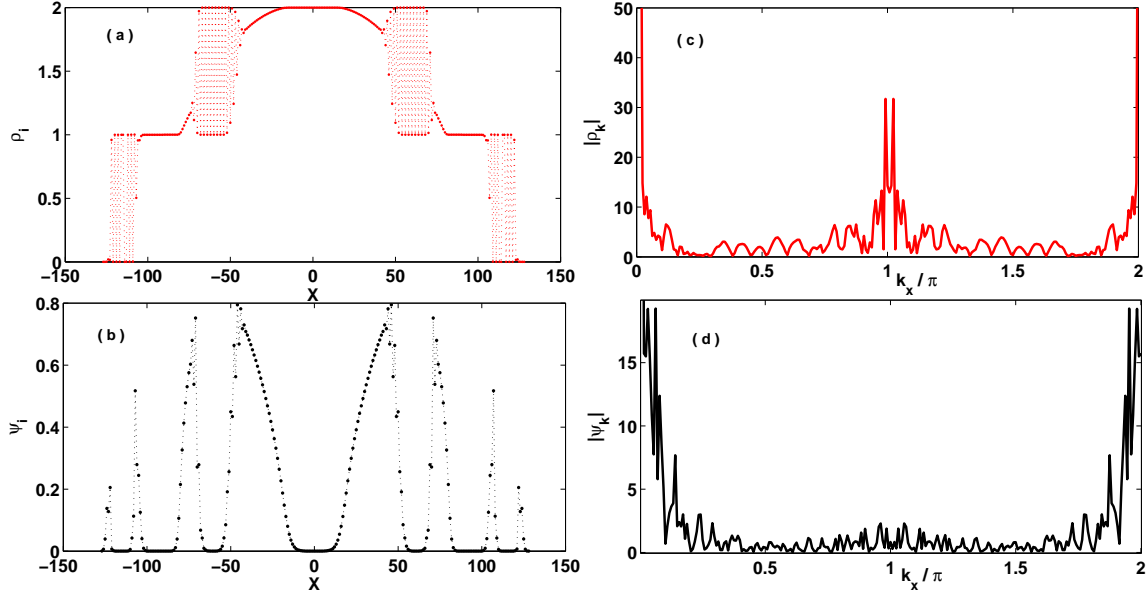,width=\linewidth,height=8cm}
\caption{(Color online) Plots of
(a) $\rho_{i}$ (red dashed line and points) and (b)
$\psi_{i}$ (black dashed line and points) versus the position $X$ along the line
$Y=Z=0$ for $V_T/(zt)=0.002,\, U/(zt)=12$ , $V/U=0.6$ and
$\mu/(zt)=30$; the moduli of the one-dimensional Fourier
transforms, namely, $|\rho_k|$ and $|\psi_k|$, of the
plots in (a) and (b) are plotted, respectively, in (c) and (d)
versus the wave vector $k_X/\pi$.}
\label{fig:rhopsi1}
\end{figure}

\begin{figure}[htbp]
\epsfig{file=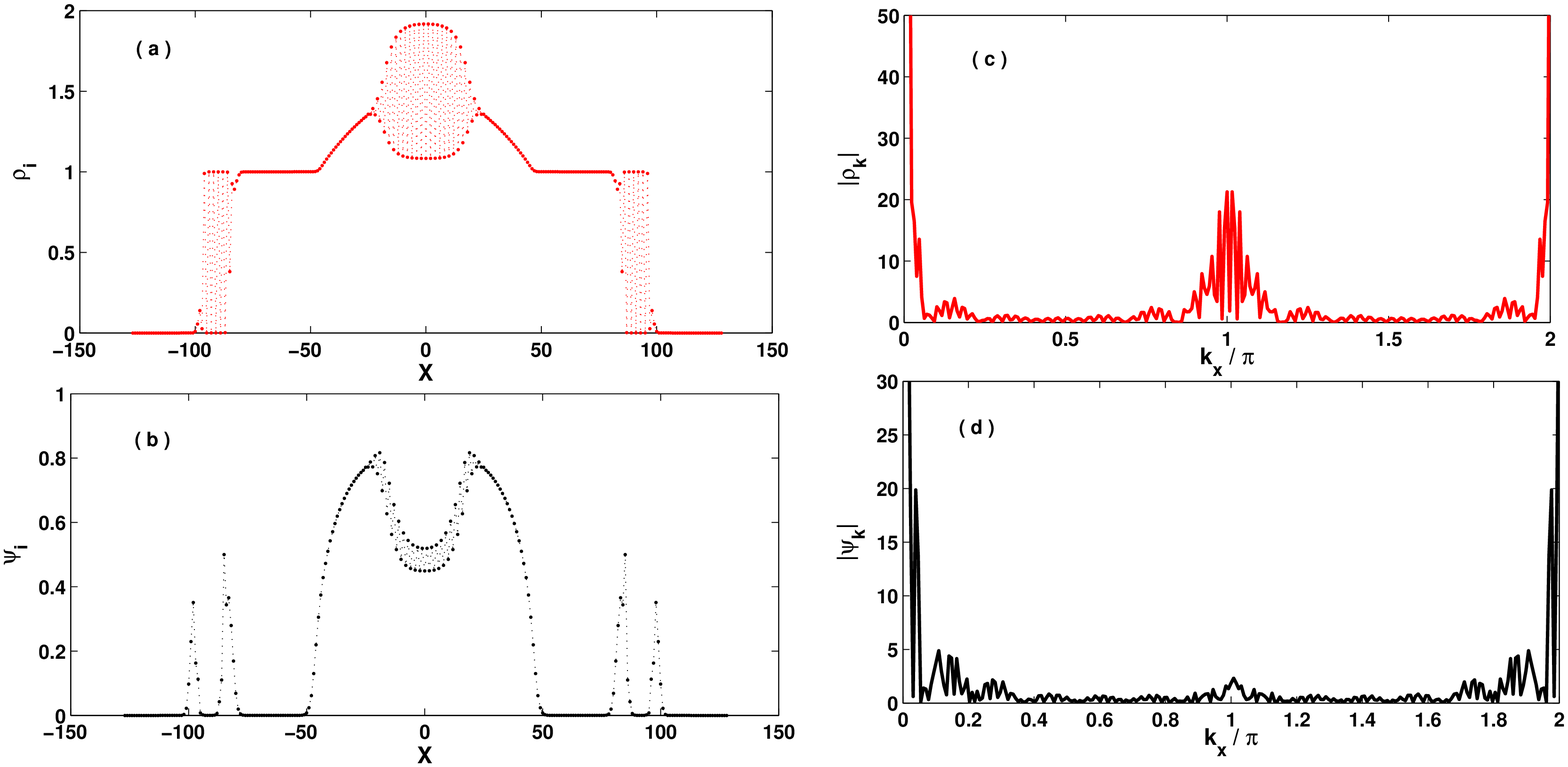,width=\linewidth,height=8cm}
\caption{(Color online)
Plots of (a) $\rho_{i}$ (red dashed line and points) and (b)
$\psi_{i}$ (black dashed line and points) versus the position $X$ along the line
$Y=Z=0$ for $V_T/(zt)=0.002,\, U/(zt)=12$ , $V/U=0.4$ and
$\mu/(zt)=19$; the moduli of the one-dimensional Fourier
transforms, namely, $|\rho_k|$ and $|\psi_k|$, of the
plots in (a) and (b) are plotted, respectively, in (c) and (d)
versus the wave vector $k_X/\pi$.}
\label{fig:rhopsi2}
\end{figure}

\begin{figure}[htbp]
\epsfig{file=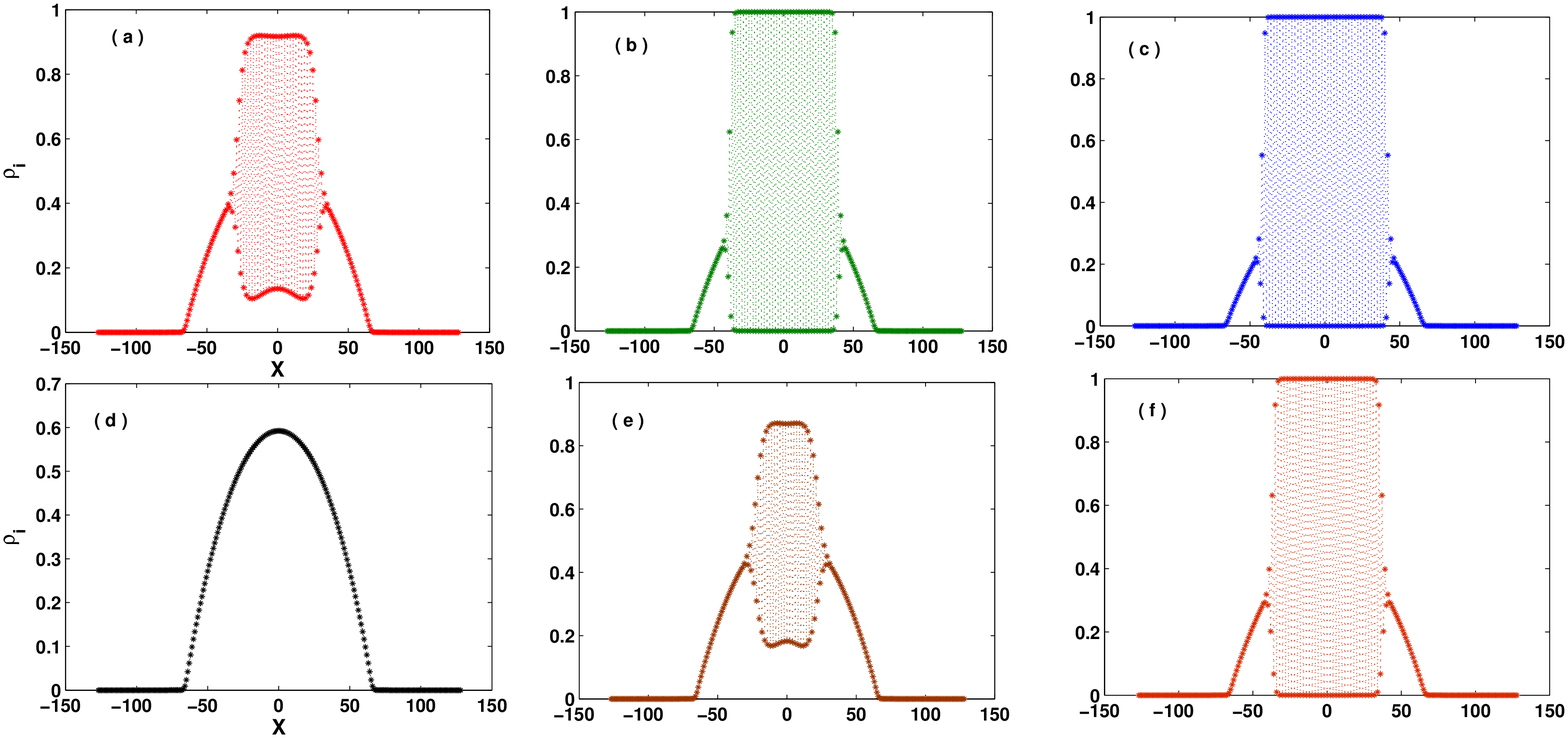,width=\linewidth,height=8cm}
\caption{(Color online) Plots of $\rho_{i}$
versus $X$, along the line $Y=Z=0$,  with $\mu/(zt) = 1.2$,
$V_T/(zt)=0.0005$, and (a) $U/(zt)=4.2$ and $V/U=0.6$,
(b) $U/(zt)=5.2$ and $V/U=0.6$, (c) $U/(zt)=6.2$ and $V/U=0.6$,
(d) $U/(zt)=4.9$ and $V/U=0.4$,  (e) $U/(zt)=5.9$ and $V/U=0.4$,
and (f) $U/(zt)=6.9$ and $V/U=0.4$.}
\label{fig:rho3}
\end{figure}
\begin{figure}[htbp]
\epsfig{file=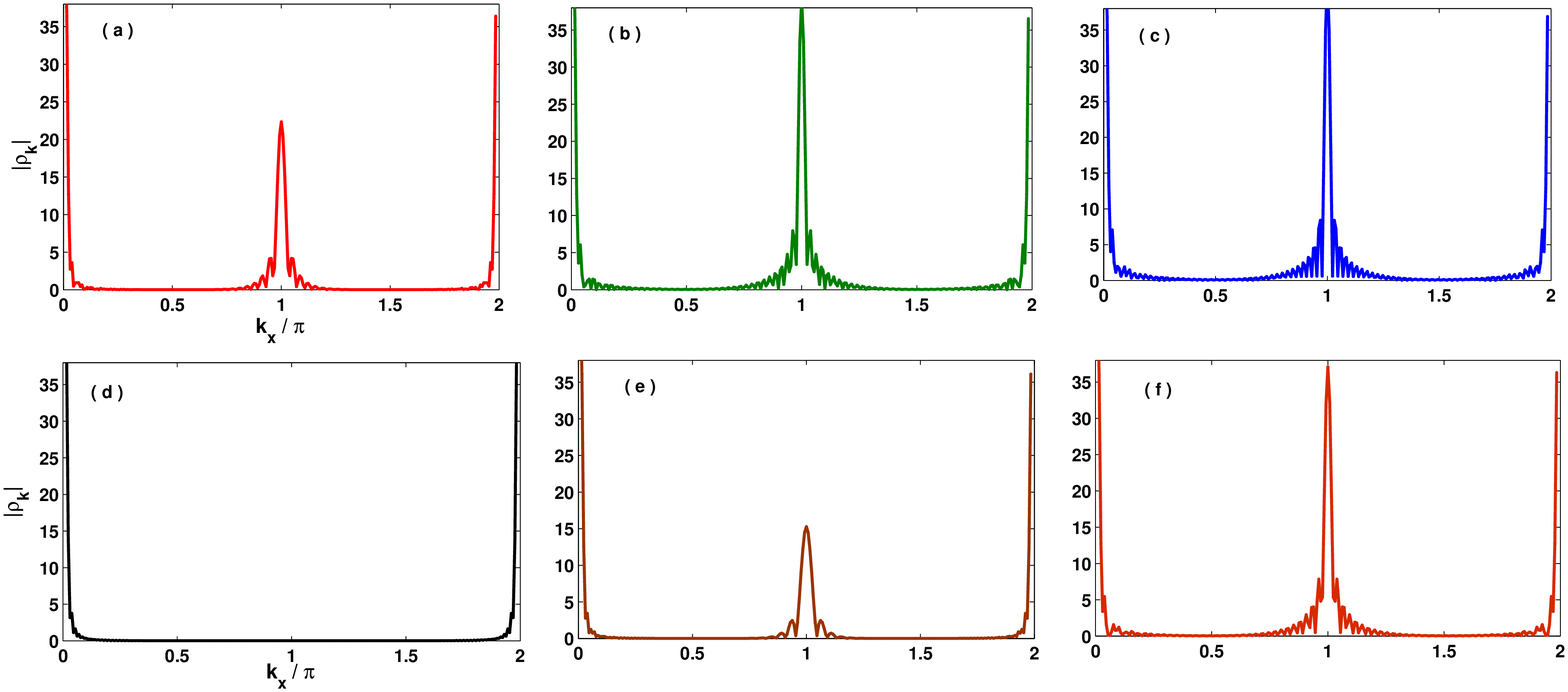,width=\linewidth,height=8cm}
\caption{(Color online) The moduli of the one-dimensional Fourier
transforms, namely, $|\rho_k|$, of the plots of $\rho_{i}$ in
Figs.~\ref{fig:rho3}(a), (b), (c), (d), (e), and (f) are plotted,
respectively, in (a),(b), (c), (d), (e), and (f) here versus the
wave vector $k_X/\pi$.}
\label{fig:rho4}
\end{figure}
\begin{figure}[htbp]
\epsfig{file=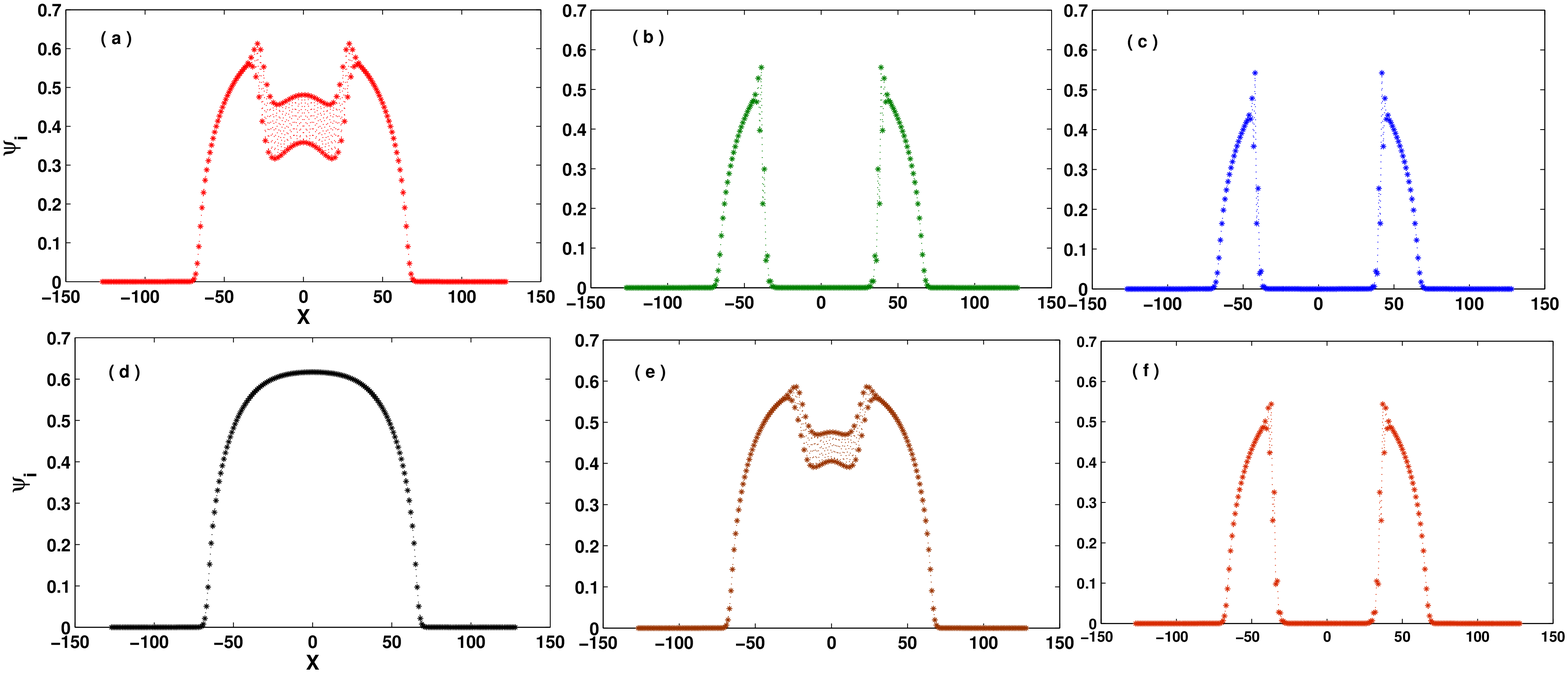,width=\linewidth,height=8cm}
\caption{(Color online) Plots of $\psi_{i}$
versus $X$, along the line $Y=Z=0$,  with $\mu/(zt) = 1.2$,
$V_T/(zt)=0.0005$, and (a) $U/(zt)=4.2$ and $V/U=0.6$,
(b) $U/(zt)=5.2$ and $V/U=0.6$, (c) $U/(zt)=6.2$ and $V/U=0.6$,
(d) $U/(zt)=4.9$ and $V/U=0.4$,  (e) $U/(zt)=5.9$ and $V/U=0.4$,
and (f) $U/(zt)=6.9$ and $V/U=0.4$.}
\label{fig:psi3}
\end{figure}
\begin{figure}[htbp]
\epsfig{file=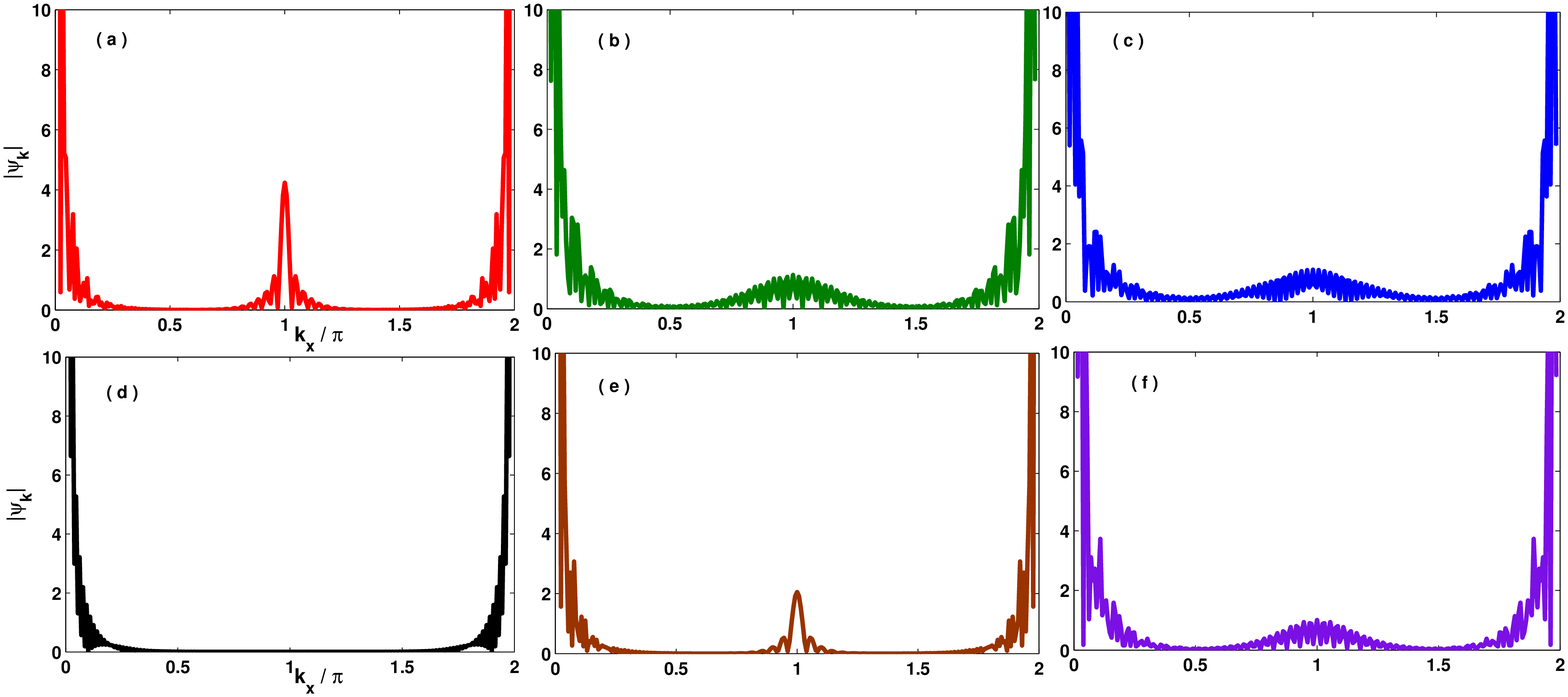,width=\linewidth,height=8cm}
\caption{(Color online) The moduli of the one-dimensional Fourier
transforms, namely, $|\psi_k|$, of the plots of $\psi_{i}$ in
Figs.~\ref{fig:psi3}(a), (b), (c), (d), (e), and (f) are plotted,
respectively, in (a),(b), (c), (d), (e), and (f) here versus the
wave vector $k_X/\pi$.}
\label{fig:psi4}
\end{figure}

In particular, we use a 3D simple-cubic lattice with $256^3$
sites and $V_T/(zt)=0.002$; and we study the following two
representative case: (a) $\mu/(zt)= 30$ and $V/U=0.6$; and (b)
$\mu/(zt)=19$ and $V/U=0.4$. With these parameters the total
number of bosons $N_T\simeq 10^6$, which is comparable to
experimental values.  Furthermore, this choice of parameters
leads not only to SF shells and two well-developed MI shells (MI1
and MI2) but also to two well-developed DW shells (DW 1/2 and DW
3/2) and SS shells.

Before we study this shell structure let us explore some
order-parameter profiles.  Plots of the order parameters
$\rho_{i}$ (red dashed line and points) and $\psi_{i}$ (black
dashed line and points) versus the position $X$ along the line $Y=Z=0$
 are shown in Figs.~\ref{fig:rhopsi1} (a) and (b), respectively,
for $V_T/(zt)=0.002,\, U/(zt)=12$ , $V/U=0.6$, and $\mu/(zt)=30$.
These plots show that the region near $X=0$ is an MI2 phase
with $\rho_i = 2$ and $\psi_i =0$.  As we move outwards from here
(either towards $X>0$ or $X<0$), we emerge into an SF phase with a
noninteger value of $\rho_i$ and $\psi_i > 0$; note that $\rho_i$
and $\psi_i$ do not oscillate here as functions of $X$. At
slightly larger values of $|X|$ the system moves into a very
narrow SS region in which both $\rho_i$ and $\psi_i$ are
oscillating functions of $X$. If we increase $|X|$, this SS
phase gives way to a DW 3/2 phase in which $\rho_i$ oscillates as
a function of $X$ but $\psi_i = 0$.  A further increase in $|X|$
yields another very narrow region of the SS phase; this is then
followed by a narrow SF region.  As we increase $|X|$ some more,
the MI1 phase is stabilized; this is followed by a very narrow SF
region, which is, in turn, followed by a narrow SS region, and
then a DW 1/2 regime. This gives way to a very narrow SS
region, as $|X|$ increases even more; finally we enter a small
region in which the boson density vanishes. Such profiles of
$\rho_i$ and $\psi_i$ imply the shell structure that we explore below.

It is also useful to obtain a complementary,
Fourier-representation picture of the profiles in
Figs.~\ref{fig:rhopsi1} (a) and (b), because it might be possible to
obtain them in time-of-flight measurements (see, e.g., Eq.~(44)
in Ref.~\cite{rmp}). Three-dimensional transforms of the shell
structure can be obtained, but they are not easy to visualize;
therefore, we present the one-dimensional Fourier transforms of
$\rho_i(X,Y=0,Z=0)$ and $\psi_i(X,Y=0,Z=0)$ with respect to $X$.
The moduli of these transforms, namely, $|\rho_k|$ and
$|\psi_k|$, of the profiles in Figs.~\ref{fig:rhopsi1} (a) and (b)
are plotted, respectively, in Figs.~\ref{fig:rhopsi1} (c) and (d)
versus the wave vector $k_X/\pi$. The principal peaks in these
transforms occur at $k_X = 0$ (or $2 \pi$) and $k_X = \pi$; the
former is associated with the uniform MI and SF phases; and the
latter arises from DW and SS phases in which real-space profiles
oscillate as explained above. In an infinite system with no
confining potential, these are the only peaks; however, as we
have seen above, the quadratic confining potential leads to
shells of MI, SF, SS, and DW phases; this shell structure leads
to the subsidiary peaks that appear in  Figs.~\ref{fig:rhopsi1} (c)
and (d) away from $k_X = 0, \, \pi$ and $2 \pi$.

Analogs of the order-parameter profiles of
Figs.~\ref{fig:rhopsi1} (a) and (b) are given in
Figs.~\ref{fig:rhopsi2} (a) and (b), for   $\rho_{i}$ (red dashed
line and points) and $\psi_{i}$ (black dashed line and points),
respectively, versus the position $X$ along the line $Y=Z=0$ for
$V_T/(zt)=0.002,\, U/(zt)=12$ , $V/U=0.4$ and $\mu/(zt)=19$. From
these plots we see that, in this case, the sequence of phases is
SS, SF, MI1, SF, a narrow strip of SS, then DW 1/2, another
narrow sliver of SS, and finally a region with vanishing boson
density. The moduli of the one-dimensional Fourier transforms,
namely, $|\rho_k|$ and $|\psi_k|$, of the plots in
Figs.~\ref{fig:rhopsi2} (a) and (b) are plotted, respectively, in
Figs.~\ref{fig:rhopsi2} (c) and (d) versus the wave vector
$k_X/\pi$.

From the profiles in Figs.~\ref{fig:rhopsi1} (a) and (b) and
Figs.~\ref{fig:rhopsi2} (a) and (b) it is clear that the precise sequence of
MI, SF, SS, and DW shells
depends on the parameters in the extended Bose-Hubbard
model~(\ref{eq:ebhmodel}). We illustrate this for
other sets of parameter values via representative
plots, in Figs.~\ref{fig:rho3} (a)-(f), of the density order
parameter $\rho_{i}$ versus $X$, along the line $Y=Z=0$,  with
$\mu/(zt) = 1.2$, $V_T/(zt)=0.0005$, and (a) $U/(zt)=4.2$ and
$V/U=0.6$,  (b) $U/(zt)=5.2$ and $V/U=0.6$, (c) $U/(zt)=6.2$ and
$V/U=0.6$, (d) $U/(zt)=4.9$ and $V/U=0.4$,  (e) $U/(zt)=5.9$ and
$V/U=0.4$, and (f) $U/(zt)=6.9$ and $V/U=0.4$, respectively. The
moduli of the one-dimensional Fourier transforms, namely,
$|\rho_k|$, of the plots of $\rho_{i}$ in Figs.~\ref{fig:rho3}
(a), (b), (c), (d), (e), and (f) are plotted, respectively, in
Figs.~\ref{fig:rho4} (a),(b), (c), (d), (e), and (f) versus the
wave vector $k_X/\pi$. The corresponding real-space plots of
$\psi_i$ and the Fourier-space plots of $|\psi_k|$ are given,
respectively, in Figs.~\ref{fig:psi3} and \ref{fig:psi4}.

\begin{figure}[htbp]
\centering \epsfig{file=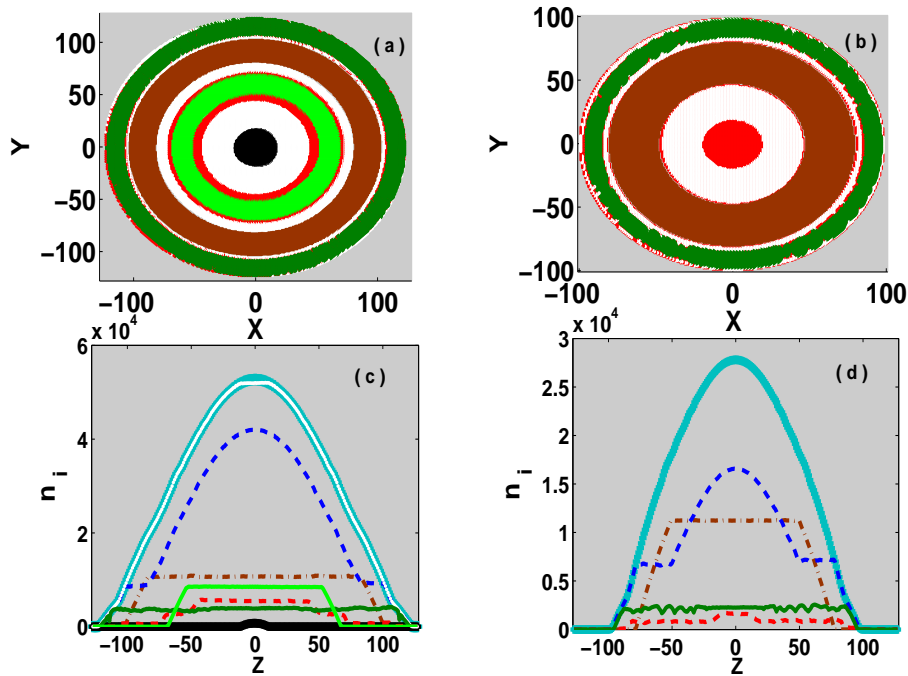,width=\linewidth,height=9cm}
\caption{(Color online) SF (white), MI2 (black), MI1 (brown), SS
(red), DW 3/2 (light green), and DW 1/2 (dark green) annuli in
the 2D planar section  ${\cal P}_{Z=0}$  (see text) with $U/(zt)
= 12$ and $V_T/(zt)=0.002$ and (a) $V/U = 0.6$ and $\mu/(zt)= 30$
and (b) $V/U = 0.4$ and $\mu/(zt)= 19$; for these parameter
values, the  integrated, in-trap density profiles are plotted
versus $Z$ in (c) and (d), respectively. These in-trap profiles
show the total number of bosons $N_T$ (light blue full lines),
the number of bosons in MI2 and MI1 regions, $N_2$ (black line in
(c)) and $N_1$ (brown dash-dotted lines), respectively, the
numbers of bosons in DW 3/2 (light green line in (c)) and DW 1/2
(dark green line in (d)) regions,  the numbers of bosons in SS
regions (red dashed line), [$N_{T}-N_{2}$] (white full line in
(c)), and [$N_{T}-N_{1}$] (blue dashed lines) in (c) and (d).  The
outermost gray regions contain no bosons.}
\label{fig:shells}
\end{figure}

These SF, MI, DW, and SS shells appear as annuli~\cite{rvpaiinhom11}
in a two-dimensional (2D) planar section ${\cal P}_Z$ through the 3D
lattice, at a vertical distance $Z$ from the center as shown, for
${\cal P}_{Z=0}$, $U/(zt)=12$, and $V_T/(zt) = 0.002$, in
Fig.~\ref{fig:shells} (a), with $ V/U=0.6$ and $\mu/(zt)=30$,  and
Fig.~\ref{fig:shells} (b), with $V/U=0.4$ and  $\mu/(zt)=19$. In the
former case, the core region near $X=Y=Z=0$, has an MI2 phase,
whereas, in the latter case, this central region is an SS phase. As
we move radially outward from the center, shells of other phases
appear; the sequence of shells in Fig.~\ref{fig:shells} (a) is the
one that results from the order-parameter profiles in
Figs.~\ref{fig:rhopsi1} (a) and (b); and the sequence of shells in
Fig.~\ref{fig:shells} (b) follows from the profiles in
Figs.~\ref{fig:rhopsi2} (a) and (b).

For any 2D planar section ${\cal P}_Z$ we can calculate
integrated, in-trap density profiles such as $N_m(Z)$, the number
of bosons in the $\rho=m$ MI annuli; similarly, we can calculate
$N_{p/q}(Z)$ in the $\rho=p/q$ DW annuli. [Here $m, \, p$, and
$q$ are intergers; e.g., we study $m=1$ or $m=2$ and $p/q =1/2$
and $p/q = 3/2$.]  We can also calculate the remaining number of
bosons, e.g., $N^r_m(Z)=N_{T}-N_m(Z)$.  For the parameter values
of Figs.~\ref{fig:shells} (a) and (b), illustrative integrated,
in-trap density profiles are plotted versus $Z$ in
Figs.~\ref{fig:shells} (c) and (d), respectively. These in-trap
profiles show the total number of bosons $N_T$ (light blue full
lines), the number of bosons in MI2 and MI1 regions, $N_2$ (black
line in Fig.~\ref{fig:shells} (c)) and $N_1$ (brown dash-dotted
lines), respectively, the numbers of bosons in DW 3/2 (light
green line in Fig.~\ref{fig:shells} (c)) and DW 1/2 (dark green
line in  Fig.~\ref{fig:shells} (d)) regions,  the numbers of
bosons in SS regions (red dashed line), [$N_{T}-N_{2}$] (white
full line in Fig.~\ref{fig:shells} (c)), and [$N_{T}-N_{1}$]
(blue dashed lines).  The outermost gray regions contain no
bosons. Such integrated, in-trap density profiles have been
obtained experimentally in Ref.~\cite{bloch} (see, e.g.,
their Fig. (3)) for cold-atom systems with SF and MI phases;
therefore, it should be possible to carry out similar experiments
on the dipolar systems~\cite{werner05,goral02} that have
motivated our study.

\section{Conclusions}
\label{sec:conclusions}

We have developed an inhomogeneous mean-field theory for the phases and
order-parameter profiles of the inhomogeneous, extended Bose-Hubbard
model~(\ref{eq:ebhmodel}) by generalizing earlier studies for the
spinless~\cite{sheshadri,rvpaiinhom11} and spin-1~\cite{rvpai08} Bose-Hubbard
models. In the homogeneous case, our theory leads to SF, MI, DW, and SS
phases and phase diagrams, such as those of Figs.~\ref{fig:homphasediag} (a)
and (b); these are qualitatively similar to those obtained by a Gutzwiller
approximation in Ref.~\cite{kovrizhin05}. In the inhomogeneous case, i.e.,
with $V_T > 0$ in the extended Bose-Hubbard model~(\ref{eq:ebhmodel}), our
theory lead to rich, order parameter-profiles (see, e.g.,
Figs.~\ref{fig:rhopsi1}, \ref{fig:rhopsi2}, \ref{fig:rho3}, and
\ref{fig:psi3}). We have also explored the Fourier-space manifestations of
these profiles, the structures of the shells of SF, MI, DW, and SS phases and
the associated integrated, in-trap density profiles for representative
parameter values. Such shell structure has been explored for cold-atom
systems that can be modelled by Bose-Hubbard
models~\cite{bloch,rvpaiinhom11,bergkvist,
pollet,smita,mitra,spiel,nandini,ozaki} but not for the extended Bose-Hubbard
model.

To make a detailed comparison of our results with experiments, the
parameters of the Bose-Hubbard model must be related to experimental
ones~\cite{rmp} as follows: $\frac{U}{zt}=\frac{\sqrt{8}\pi}{4z}
\frac{a_s}{a}\exp({2\sqrt{\frac{V_0}{E_r}}})$, where $E_r$ is the
recoil energy, $V_0$ the strength of the lattice potential, $a_s$
($=5.45$~nm for $^{87}$Rb) the $s$-wave scattering coefficient,
$a=\lambda/2$ the optical lattice constant, and $\lambda=825$~nm the
wavelength of the laser used to create the optical lattice;
typically $0 \leq V_0 \leq 22 E_r$. If we use this experimental
parametrization, we scale all the energies by $E_r$. [In this paper,
we set $zt=1$, i.e., we measure all energies in units of $zt $.] For
the extended Bose-Hubbard case, the relation of our model parameters
to parameters in dipolar systems~\cite{werner05,goral02} is not
straightforward because of the long-range interactions. However,
rough estimates can be made as follows:
\begin{equation}\label{eq:t}
    t=\int
    w^*({\bf r}-{\bf r}_i)[\frac{-\hbar^2}{2m}{\nabla^2}+V_l({\bf r})]w({\bf r}-{\bf r}_j)d^3r ,
\end{equation}
where $i$ and $j$ are nearest-neighbor sites, $w$ are Wannier
functions, and $ V_l({\bf r})=\sum_{\alpha=x,y,z}V^2_\alpha
\cos^2(k_\alpha \alpha)$ is the optical-lattice potential with
wavevector $\bf{k}$. Furthermore,

\begin{equation}\label{eq:U}
U=U_{ii}=\int |w({\bf r}- {\bf r}_i)|^2 V_{\rm int}({\bf r}-{\bf
r'}) |w({\bf r'}- {\bf r}_i)|^2 d^3r~ d^3{r'}
\end{equation}
and
\begin{equation}\label{eq:V}
V=U_{<ij>}=\int |w({\bf r}- {\bf r}_i)|^2 V_{\rm int}({\bf r}-{\bf
r'}) |w({\bf r'}- {\bf r}_j)|^2 d^3r~ d^3{r'},
\end{equation}
with
\begin{equation}\label{eq:VINT}
V_{\rm int}=D^2 \frac{1-3\cos^2\theta}{|{\bf r}-{\bf r'}|^3} +
\frac{4\pi\hbar^2a_s}{m}\delta({\bf r}-{\bf r'}).
\end{equation}
Here $D$ is the dipole moment, $a_s$ is the $s$-wave  scattering
constant, and $m$ is the mass. The $s$-wave scattering constant of
chromium is $|a(^{52}{\rm Cr})|=(170\pm39)a_0$ and $|a(^{50}{\rm
Cr})|=(40\pm15)a_0$, where $a_0=0.053$~nm~\cite{chrom03}.

We hope that our work will stimulate experiments designed to explore SF, MI,
DW, and SS shells in dipolar-condensate systems~\cite{werner05,goral02}.

\section{Acknowledgments}
\label{sec:acknowl}

We thank A. Bhatnagar, K. Rajany, and especially K. Sheshadri , S. Mukerjee and S. Bhattacharjee for discussions,
and DST, CSIR, and UGC (India) for support. We  would like to
dedicate this paper to Professor Ulrich Eckern on the occasion of
his sixtieth birthday.


\begin{thebibliography}{11}

\bibitem{rmp} I. Bloch, J. Dalibard, and W. Zwerger, Rev. Mod. Phys.
\textbf{80}, 885 (2008).

\bibitem{advphys}M. Lewenstein, {\it et al.}, Adv. in Physics,
\textbf{56} 243 (2007).

\bibitem{jaksch} D. Jaksch, {\it et al.}, Phys. Rev. Lett.
\textbf{81}, 3108 (1998).

\bibitem{greiner} M. Greiner, {\it et al.}, Nature (London)
\textbf{415}, 39 (2002).

\bibitem{fisher} M.P.A. Fisher, {\it et al.}, Phys. Rev. B
\textbf{40}, 546 (1989); D.S. Rokhsar and B. G. Kotliar, Phys. Rev. B
\textbf{44}, 10328 (1991); W. Krauth, M. Caffarel, and J-P. Bouchaud,
Phys. Rev. B \textbf{45}, 3137 (1992).

\bibitem{mc}W. Krauth, N. Trivedi, and D. Ceperley, Phys. Rev. Lett.
\textbf{67}, 2307 (1991); N. Trivedi and M. Makivic, \emph{ibid.}
\textbf{74}, 1039 (1995).

\bibitem{sheshadri} K. Sheshadri, {\it et al.}, Europhys. Lett. \textbf{22}
257 (1993).


\bibitem{wessel} S. Wessel, {\it et al.}, Phys. Rev. A, \textbf{70} 053615
(2004).

\bibitem{svistunov} V.A. Kashurnikov, N.V. Prokofev, and B.V.
Svistunov, Phys. Rev. A, \textbf{66}, 031601 (2002).

\bibitem{bloch} S. F\"olling, {\it et al.}, Phys. Rev. Lett.  \textbf{97},
060403 (2006).

\bibitem{campbel} G.K. Campbell, {\it et al.}, Science \textbf{313}, 649
(2006).

\bibitem{rvpai08} R. V. Pai, K. Sheshadri and R. Pandit, Phys. Rev.  B
\textbf{77}, 014503 (2008).

\bibitem{sheshprl} K. Sheshadri, {\it et al.}, Phys. Rev. Lett.  \textbf{75}
4075 (1995).

\bibitem{rvpaiinhom11} R.V. Pai, J.M. Kurdestany, K. Sheshadri,
and R. Pandit, arXiv:1201.1642 (2012).

\bibitem{kovrizhin05} D. Kovrizhin, G.V. Pai, S. Sinha, Europhys.
Lett. \textbf{72}, 162 (2005).

\bibitem{werner05} J. Werner, {\it et al.}, Phys. Rev. Lett.
\textbf{94}, 183201 (2005).

\bibitem{goral02} K. G\"oral, L. Santos, M. Lewenstein, Phys.
Rev. Lett. \textbf{88}, 170406 (2002).

\bibitem{andreev69} A.F. Andreev and I.M. Lifshitz, Sov. Phys.
JETP, \textbf{29}, 1107 (1969); A.J. Leggett, Phys.
Rev. Lett. \textbf{25}, 1543 (1970); G. Chester, Phys. Rev. A \textbf{2},
256 (1970).

\bibitem{kim04} E. Kim and M.H.W. Chan, Nature \textbf{427}, 225 (2004).



\bibitem{bergkvist}S. Bergkvist, P. Henelius, and A.  Rosengren, Phys. Rev. A
\textbf{70}, 053601 (2004).

\bibitem{pollet}L. Pollet, {\it et al.}, Phys. Rev. A \textbf{69}, 043601
(2004).

\bibitem{smita}B. DeMarco, {\it et al.}, Phys.  Rev. A \textbf{71}, 063601
(2005).

\bibitem{mitra}K. Mitra, C.J. Williams, and C. A. R. S\'a de Melo,
Phys. Rev. A \textbf{77}, 033607 (2008).

\bibitem{spiel} I. B. Spielman, W. D. Phillips, and J. V. Porto, Phys. Rev.
Lett. \textbf{98}, 080404 (2007); {\it{ibid}} \textbf{100}, 120402 (2008).

\bibitem{nandini} Y. Kato, {\it et al.}, Nature Physics  \textbf{4}, 617 (2008).

\bibitem{ozaki} T. Ozaki and T. Nikuni, J. Phys. Conference
Series \textbf{150}, 042158 (2009).

\bibitem{chrom03} P.O. Schmidt, {\it et al.}, Phys. Rev. Lett. \textbf{91},
193201 (2003).

\end{thebibliography}
\end{document}